\documentclass[aps,prl,reprint,groupedaddress,superscriptaddress,longbibliography]{revtex4-1}
\usepackage{mathrsfs}
\usepackage{blindtext}
\usepackage{soul}
\usepackage{amssymb,amsbsy,amsmath,latexsym,dsfont,array,layout,graphicx,mathrsfs,color,ulem,bm}

\newcommand{\one}{\mathds{1}}
\newcommand{\ket}[1]{\left|{#1}\right\rangle}
\newcommand{\bra}[1]{\left\langle{#1}\right|}

\newcommand{\targetstate}{\Phi_{A'B'}}
\newcommand{\tr}{\operatorname{Tr}}

\begin{document}

\title{Experimental demonstration of one-sided device-independent self-testing of any pure two-qubit entangled state}
\author{Zhihao Bian}
\affiliation{Beijing Computational Science Research Center, Beijing 100084, China}
\affiliation{School of Science, Jiangnan University, Wuxi 214122, China}
\author{A. S. Majumdar}
\affiliation{S. N. Bose National Centre for Basic Sciences, Salt Lake, Kolkata 700 106, India}
\author{C. Jebaratnam}
\affiliation{S. N. Bose National Centre for Basic Sciences, Salt Lake, Kolkata 700 106, India}
\affiliation{Department of Physics and Center for Quantum Frontiers of Research and Technology (QFort), National Cheng Kung University, Tainan 701, Taiwan}
\author{Kunkun Wang}
\affiliation{Beijing Computational Science Research Center, Beijing 100084, China}
\author{Lei Xiao}
\affiliation{Beijing Computational Science Research Center, Beijing 100084, China}
\affiliation{Department of Physics, Southeast University, Nanjing 211189, China}
\author{Xiang Zhan}
\affiliation{Beijing Computational Science Research Center, Beijing 100084, China}
\affiliation{School of Science, Nanjing University of Science and Technology, Nanjing 210094, China}
\author{Yongsheng Zhang}
\affiliation{Key Laboratory of Quantum Information, University of Science and Technology of China, CAS, Hefei 230026, China}
\affiliation{Synergetic Innovation Center in Quantum Information and Quantum Physics, University of Science and Technology of China, CAS, Hefei 230026, China}
\author{Peng Xue}\email{gnep.eux@gmail.com}
\affiliation{Beijing Computational Science Research Center, Beijing 100084, China}

\begin{abstract}
We demonstrate one-sided device-independent self-testing of any pure entangled two-qubit state based on a fine-grained steering inequality. The maximum violation of a fine-grained steering inequality can be used to witness certain steerable correlations, which certify all pure two-qubit entangled states. Our experimental results identify which particular pure two-qubit entangled state has been self-tested and which measurement operators are used on the untrusted side. Furthermore, we analytically derive the robustness bound of our protocol, enabling our subsequent experimental verification of robustness through state tomography. Finally, we ensure that the requisite no-signalling constraints are maintained in the experiment.
\end{abstract}

\maketitle

%\label{sec:intro}  % \label{} allows reference to this section

{\it Introduction:}---The goal of self-testing~\cite{MY04,MYS12,YN13,CGS17} is to certify {\it a priori} unknown quantum systems in a device-independent way based on measurement statistics. Self-testing is an important aspect of quantum information~\cite{MS16}, especially with the emergence of more advanced implementations of quantum computation~\cite{MM11,GWK17,RUV13} and secure communication~\cite{BCW+12,ABG+07,MPA11}. Demonstration of quantum entanglement can certify devices to contain systems associated with particular quantum states and measurements~\cite{B64,BCP+14}. The theory of device-independent robust self-testing via Bell tests~\cite{MY04,MYS12,YN13,RZS12,BP15,K16,SAT+17,C17,WPD+18} enables several quantum states and measurements to be verified without direct access to the quantum system.  However, violation of a Bell inequality requires resources that may be difficult to implement in practice~\cite{SH16}, and is not the only method for detecting entanglement in general. On the other hand, if all devices are trusted, we have direct access to the quantum state and can do full state tomography to verify entanglement.

Between these two scenarios, there exists a third option where only one side which is trusted and we have direct access to only one part of the quantum device~\cite{GWK17,BCW+12,SH16,BVQ+14,WJD07,QVC+15,WWB+16}. Recently, one-sided device-independent self-testing (1sDIST) of any pure entangled two-qubit state based on a fine-grained steering inequality (FGSI)~\cite{PKM14} has been proposed theoretically~\cite{GBD+18}. The FGSI is derived from the fine-grained uncertainty relation~\cite{OP10} linking nonlocality with uncertainty~\cite{OP10,PM12}, which forms the basis of several information theoretic applications~\cite{PCM13,DPM13,PKM14,CPM15}. The FGSI for two-qubit states has been experimentally verified recently~\cite{OKV+18}. In the $2$-$2$-$2$ steering scenario (involving $2$ parties, $2$ measurement settings per party, $2$ outcomes per measurement setting), the maximum violation of the FGSI can be used to witness certain extremal steerable correlations, which certify all pure two-qubit entangled states~\cite{GBD+18}. This motivates the experimental study of self-testing via the FGSI, which furnishes an intermediate form of entanglement verification between full state tomography and Bell tests.

%\begin{figure}%[htbp]
%\includegraphics[width=0.5\textwidth]{fig1}
%\caption{Schematic diagram for 1sDIST of any pure two-qubit entangled state. Two spatially separated parties Alice and Bob share a quantum state which they want to self-test. Alice performs a set of black-box measurements $A_i$, while Bob performs quantum measurements $B_i$. The statistics of the outcomes of the measurements are used to verify violation of the FGSI.}
%\label{fig:scheme}
%\end{figure}

\begin{figure}%[htbp]
\includegraphics[width=0.5\textwidth]{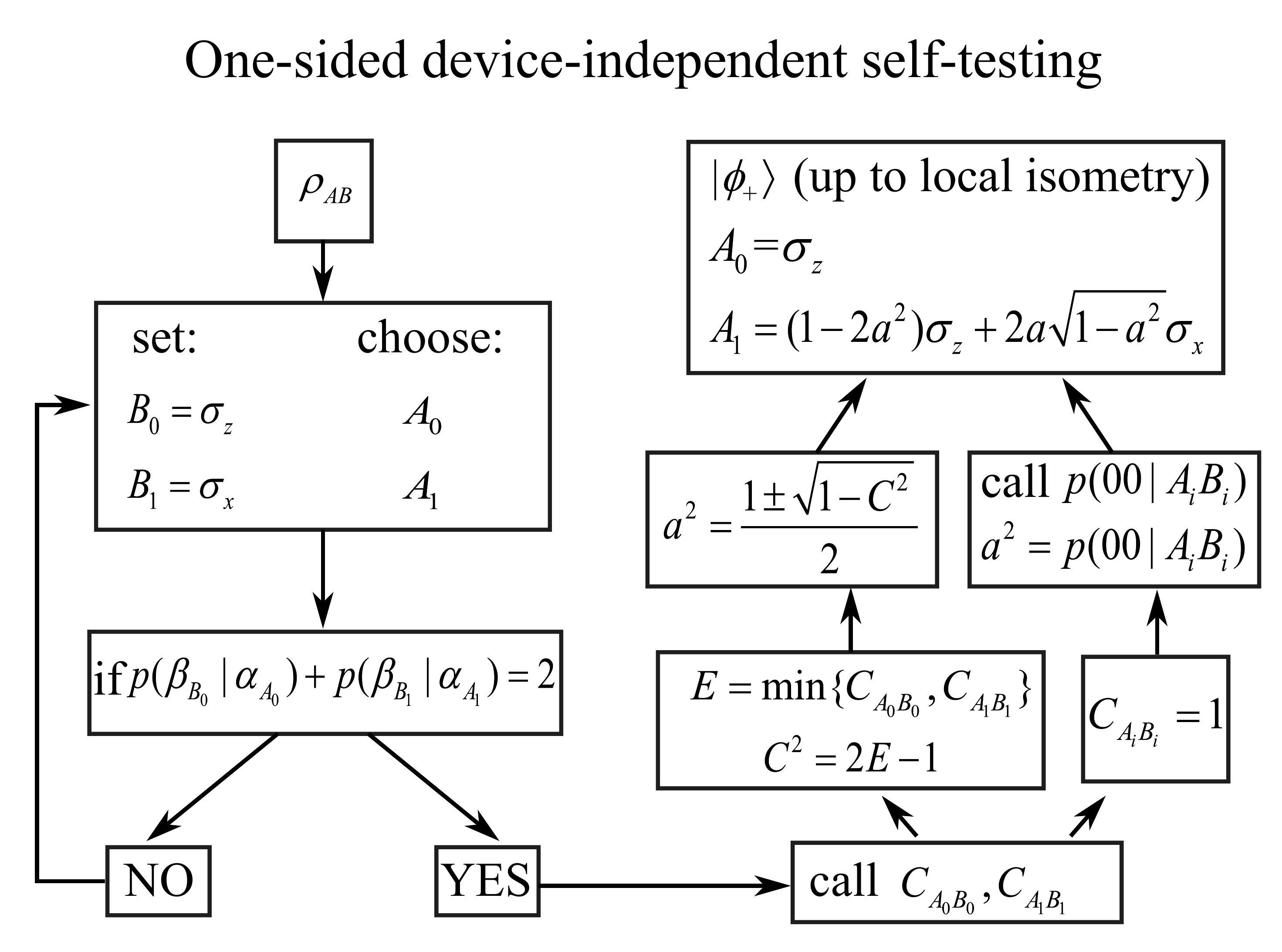}
\caption{Circuit for 1sDIST of any pure two-qubit entangled state as well as Alice's measurements.}
\label{fig:circuit}
\end{figure}

In this paper, we provide experimental evidence of 1sDIST of any pure entangled two-qubit state by using the  violation of the FGSI and a quantity called mutual predictability  which has been used for constructing entanglement witness~\cite{SHB+12} and steering inequality~\cite{LLC+18}. Our protocol can certify  which particular pure two-qubit entangled state has been self-tested, and which measurements have been performed by the untrusted party as well, based on the maximal violation of the FGSI.

In practical quantum information processing, errors are unavoidable and the robustness is essential for self-testing proposals. Here we extend the operator inequality approach~\cite{K16} to the case of our 1sDIST scheme based on the FGSI in order to analytically derive the robustness bound of our 1sDIST protocol. This enables us to show that the experimental results exhibit the required robustness by achieving excellent precision.

{\it Self-testing protocol:}---We consider a steering scenario where Alice (the untrusted party) prepares a bipartite state $\rho_{AB}$ between systems A and B, and tries to convince Bob (the trusted party) that $\rho_{AB}$ is steerable. Alice performs two black-box dichotomic measurements and Bob performs two-qubit measurements. Before sending system B to Bob, Alice knows that he will randomly choose either $\sigma_z$ or $\sigma_x$. Bob is convinced that $\rho_{AB}$ is steerable only when the FGSI~\cite{PKM14}
\begin{equation}
S_\text{FGSI}=p(\beta_{B_0}|\alpha_{A_0})+p(\beta_{B_1}|\alpha_{A_1}) \leq 1+\frac{1}{\sqrt{2}}
\end{equation}
is violated, where $p(\beta_{B_{0(1)}}|\alpha_{A_{0(1)}})$ is the probability of obtaining the outcome $\beta$ when Bob performs $B_{0(1)}$ on system B given that Alice obtains the outcome $\alpha$ by performing $A_{0(1)}$ on system A. %Alice demonstrates steerability of the state $\rho_{AB}$ to Bob only if the FGSI is violated.
The maximum violation of the FGSI is $S_\text{FGSI}=2$ if and only if $p(\beta_{B_0}|\alpha_{A_0})=p(\beta_{B_1}|\alpha_{A_1})=1$. It has been shown that the maximum violation of the FGSI self-tests an arbitrary pure two-qubit entangled state~\cite{GBD+18} in the above 1sDIST scenario. As any pure two-qubit entangled state can always be written in the form given by $\ket{\phi_+}=a\ket{00}+\sqrt{1-a^2}\ket{11}$ following Schmidt decomposition~\cite{P95,HJW93}, the maximum violation of the FGSI implies that the bipartite entangled state is the two-qubit entangled state in this form up to local isometries.

\begin{figure}%[htbp]
\includegraphics[width=0.5\textwidth]{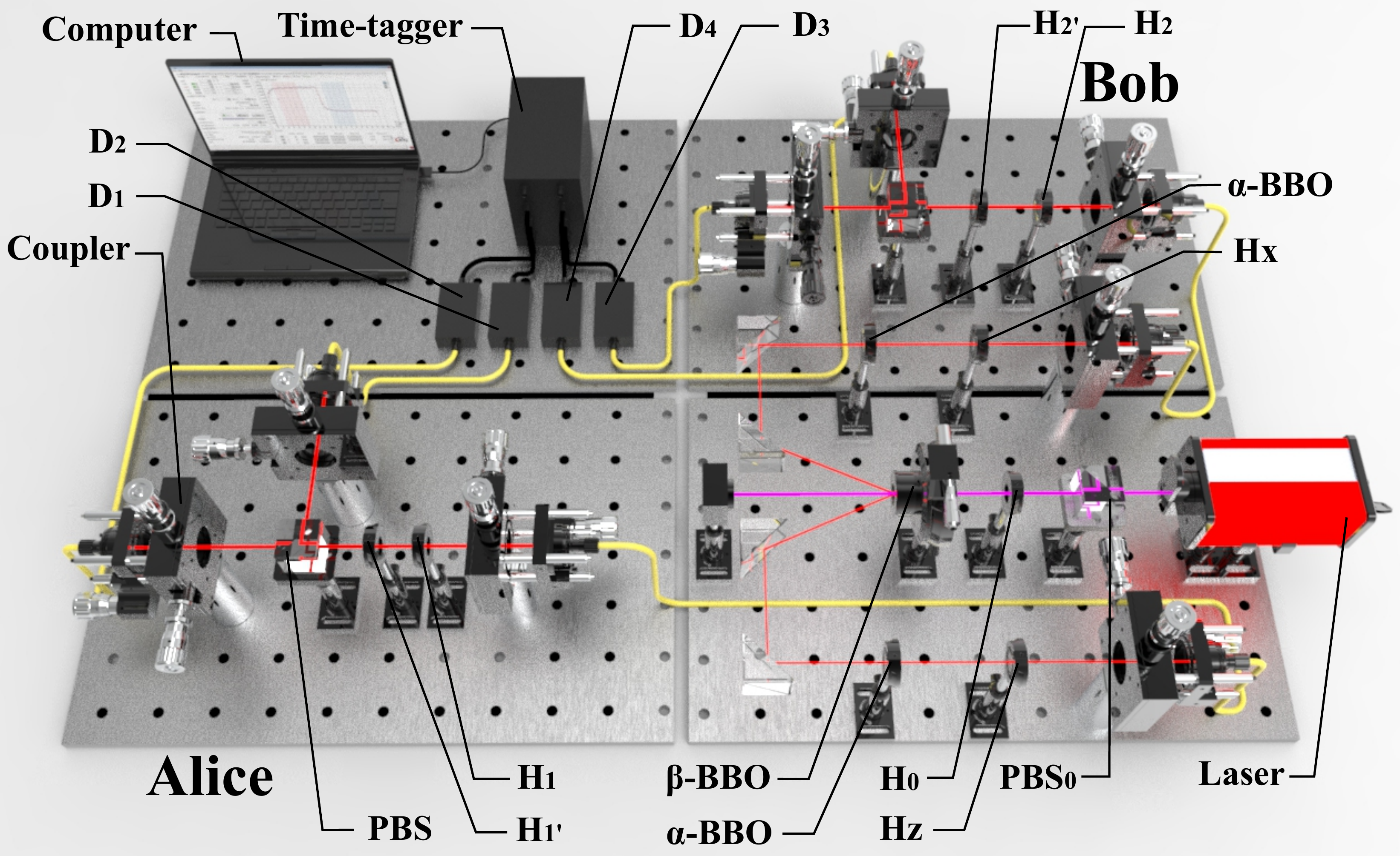}
\caption{ Experimental setup for 1sDIST of a pure two-qubit entangled state. Polarization-entangled photon pairs are generated via type-I SPDC. %where two joint $\beta$-BBO crystals are pumped by a CW diode laser and two $\alpha$-BBO crystals are used to compensate the walk-off between photons with horizontal and vertical polarizations.
For both Alice and Bob, a set of measurements can be realized by HWPs, PBS and APDs. All the joint probabilities can be read out from the coincidence between certain APDs.}
\label{fig:setup}
\end{figure}

\begin{figure*}%[htbp]
\includegraphics[width=0.8\textwidth]{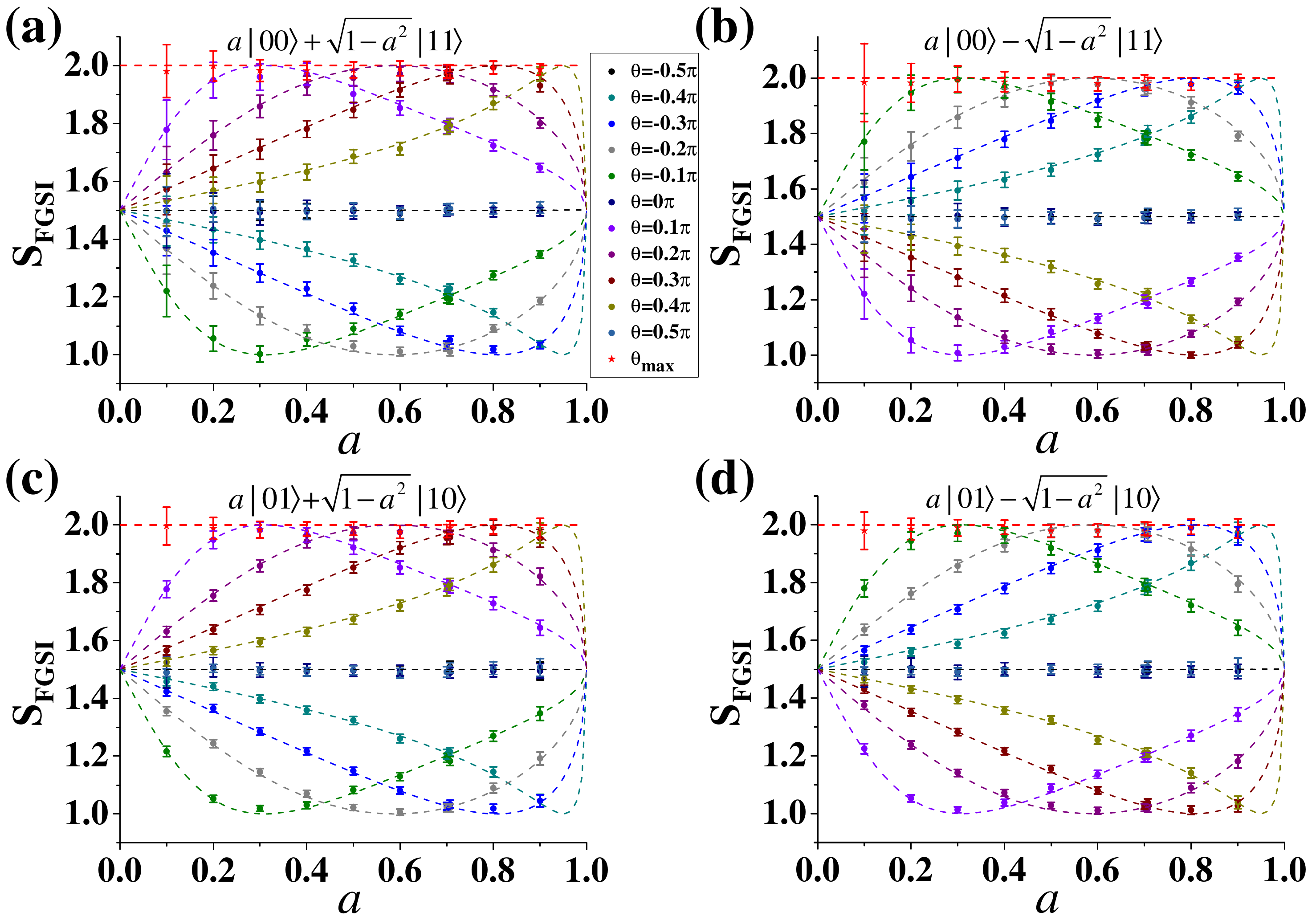}
\caption{Experimental results of $S_\text{FGSI}$ as a function of the state coefficient $a$ for different types of pure two-qubit entangled states. Different symbols represent different measurements $A_1(\theta)$ for Alice. Dashed lines represent the theoretical predictions. Error bars indicate the statistical uncertainty.}
\label{fig:FGI}
\end{figure*}

\begin{figure*}%[htbp]
\includegraphics[width=\textwidth]{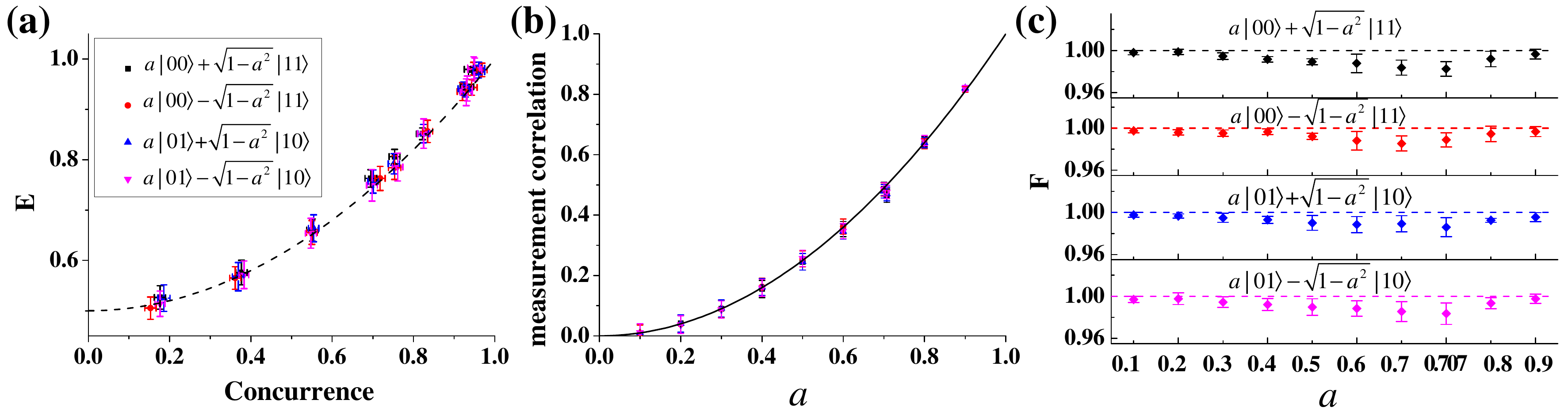}
\caption{(a) Experimental results of $E$ in Eq.~(\ref{MP}) as a function of concurrence of the state $\rho_{AB}$. (b) The measurement correlation $p(00|A_0B_0)$ as a function of the state coefficient $a$ for different four types of pure two-qubit entangled states.  %(c) Experimental results of the violation of the CFFW inequality $S_\text{CFFW}$ as a function of concurrence of the state $\rho_{AB}$. Inset: $S_\text{CFFW}$ as a function of $a$ for different types of pure two-qubit entangled states.
(c) Fidelity of the state $\ket{\phi_\pm}$ ($\ket{\psi_\pm}$) with the measured coefficient $a$ obtained via self-test and certification compared to that obtained via quantum state tomography for each initial state being self-tested.
}
\label{fig:figs}
\end{figure*}

\begin{figure*}%[htbp]
\includegraphics[width=0.8\textwidth]{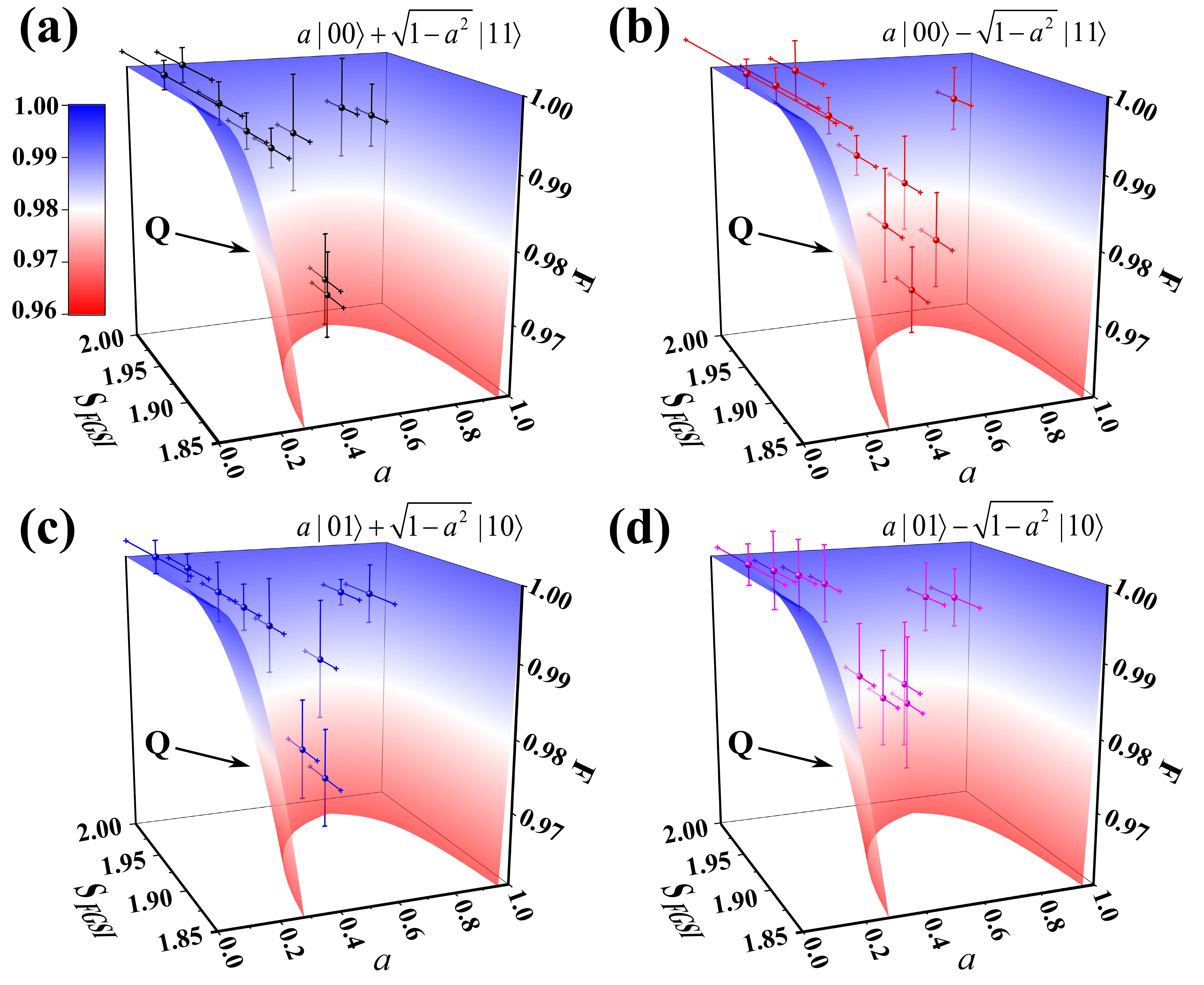}
\caption{Experimental results of fidelities between the self-tested state and the tomographically reconstructed state versus the maximum violation of the FGSI with curved surfaces representing the robustness bound $Q$ of our 1sDIST protocol. We consider the statistical noise in estimating the errors of the fidelities and the maximum violations of the FGSI. Error bars of the fidelities are obtained by a Monte Carlo simulation.}
\label{fig:robustness}
\end{figure*}

Specifically, as illustrated in Fig.~\ref{fig:circuit},
Bob fixes his observables as $B_0=\sigma_z$ and $B_1=\sigma_x$ and Alice chooses her measurements $A_0$ and $A_1$. The correlation between systems A and B violates the FGSI maximally if and only if the bipartite state is the  pure two-qubit entangled state, for instance
\begin{align}
\label{eq:state}
&\ket{\phi_\pm}=a\ket{00}\pm\sqrt{1-a^2}\ket{11},\nonumber\\
&\ket{\psi_\pm}=a\ket{01}\pm\sqrt{1-a^2}\ket{10}, 0<a<1,
\end{align}
and the choices of Alice's observables are given by $\{A_0=\sigma_z,A_1=(1-2a^2)\sigma_z\pm 2a\sqrt{1-a^2}\sigma_x\}$ for $\ket{\phi_\pm}$ and $\{A_0=-\sigma_z,A_1=(1-2a^2)\sigma_z\pm 2a\sqrt{1-a^2}\sigma_x\}$ for $\ket{\psi_\pm}$, respectively, up to local isometry. The left-hand side of the FGSI is then $S_\text{FGSI}=p(0_{B_0}|0_{A_0})+p(0_{B_1}|0_{A_1})=2$.

Here we consider the four types of states~(\ref{eq:state}) since any pure two-qubit entangled state can be transformed into one of them via local unitaries. To identify which entangled state has been self-tested, we consider the quantity $2E-1$, where
\begin{equation}
\label{MP}
E=\text{min}\{C_{A_0B_0},C_{A_1B_1}\},
\end{equation}
and
\begin{equation}
C_{A_iB_j}=\sum_{\alpha=\beta=0}^1 p(\alpha\beta|A_iB_j)
\end{equation}
which is called  mutual predictability~\cite{SHB+12} with the measurement correlation $p(\alpha\beta|A_iB_j)=\text{Tr}(\prod_{\beta|B_j}\sigma_{\alpha|A_i})$. Here $\{\prod_{\beta|B_j}\}_{\beta,B_j}$ are projective operators, $\sigma_{\alpha|A_i}=p(\alpha|A_i)\rho_{\alpha|A_i}$, $p(\alpha|A_i)$ is the conditional probability of getting the outcome $\alpha$ when Alice performs the measurement $A_i$ and $\rho_{\alpha|A_i}$ is the normalized conditional state on Bob's side. We calculate the quantity $2E-1$ for observables $A_0$ and $A_1$ acting on the Hilbert-space of dimension $d$ which implies in the circuit of Fig.~\ref{fig:circuit}
and self-testing of pure entangled states $\rho_{AB}$ acting on the Hilbert space of dimension $d \times 2$. In general, the dimension of Alice's Hilbert space is arbitrary in the context of the steering scenario that we have considered.

When the FGSI is maximally violated by any of the four pure states given by Eq.~(\ref{eq:state}), the quantity  $2E-1=C^2(\ket{\phi_\pm})=C^2(\ket{\psi_\pm})$,
where $C$ is the concurrence~\cite{CKW00}. Therefore, from the value of $2E-1$ we can infer the magnitude of entanglement of the self-tested state from the measured data that gives rise to the maximal violation of the FGSI. Note that knowing concurrence of the self-tested state also provides information about which pure state has been self-tested (up to local isometries). Moreover,  we can  determine the value of the coefficient $a$ in the pure state $\ket{\phi_+}$ that has been self-tested up to local isometry by inverting the equation $C(\ket{\phi_+})= 2a\sqrt{1-a^2}$ and obtaining $a^2=\left[1\pm\sqrt{1-C^2}\right]/2$. Thus, we can identify that Alice's measurements are $\{A_0=\sigma_z,A_1=(1-2a^2)\sigma_z+ 2a\sqrt{1-a^2}\sigma_x\}$ together with identifying which pure two-qubit entangled state in the form $\ket{\phi_+}$ has been self-tested.

Alternatively, we can identify the parameter $a$ in $\ket{\phi_+}$ directly from the measurement data. The maximal violation of the FGSI, i.e., $p(0_{A_0}|0_{B_0})+p(0_{A_1}|0_{B_1})=2$ implies that one of the mutual predictabilities $C_{A_iB_i}=1$ up to local unitaries. Suppose it turns out that  $C_{A_0B_0}=1$. Then, considering the measurement data $p(ab|00)$ corresponding to the maximal violation of the FGSI, one may deduce the value of $a$ in the pure state as follows: Note that the maximal violation of the FGSI with $C_{A_0B_0}=1$ implies the normalization $p(00|00) + p(11|00) = 1$ for any of the four pure states given by Eq.~(\ref{eq:state}). Hence, from this equation, one can determine the value of $a$ in one of the four pure states.

{\it Experimental realization:}---For experimental demonstration, a qubit is encoded by the horizontal and vertical polarizations of single photons. As illustrated in Fig.~\ref{fig:setup}, our experimental setup consists of three modules: state preparation, Alice's measurement, and Bob's measurement. In the state preparation module, entangled photons are generated via type-I spontaneous parametric down-conversion (SPDC)~\cite{PZQ+15,BLQ+15,ZZL+16,XZB+17,ZXB+17,ZKK+17,WEX+18,WQX+19,WQXZ+19,XWZ+19,Z+20}. %where two joint $0.5$mm-thick-$\beta$-barium-borate ($\beta-$BBO) crystals are pumped by a continuous-wave diode laser. %The visibility of entangled photonic state is larger than $0.950$.
%The heralded single photons pass through a polarizing beam splitter (PBS) followed by a half-wave plate (HWP, H$_0$) with the certain setting angle $\vartheta$ and the matrix from of the operation of the HWP
%$\begin{pmatrix}
%    \cos 2\vartheta & \sin 2\vartheta \\
%    \sin 2\vartheta & -\cos 2\vartheta \\
%    \end{pmatrix}$. %The matrix form of the operation of the QWP is $\begin{pmatrix} \cos^2 \vartheta+i \sin^2 \vartheta & (1-i)\sin\vartheta\cos\vartheta\\ (1-i)\sin\vartheta\cos\vartheta & \sin^2\vartheta+i\cos^2\vartheta \\ \end{pmatrix}$.
By choosing the setting angle of the half-wave plate (HWP, H$_0$) to be $\cos 2\chi=a$, photon pairs are prepared into a family of entangled state $\ket{\phi_+}=a\ket{HH}+\sqrt{1-a^2}\ket{VV}$. By inserting a HWP (H$_z$) at $0$ into one of the optical paths after the $\beta$-barium-borate (BBO) crystal, we can prepare photons into $\ket{\phi_-}$. By inserting H$_x$ at $45^\circ$ into one of the optical paths, $\ket{\psi_+}$ is generated. Whereas, by inserting both H$_z$ and H$_x$ into two optical paths respectively, we then obtain $\ket{\psi_-}$. One of the photons is sent to Bob for his measurement and the other is for Alice's measurement.

To measure Alice's observable $A_0$, a polarizing beam splitter (PBS) and two single-photon avalanche photodiodes (APDs) are used. For the measurement $A_1$, we scan the coefficients of $A_1$ by tuning the setting angles of the HWPs (H$_1$ and H$_{1'}$) following by a PBS and two APDs. Two outcomes of $A_i$ are  read by APDs (D$_1$ and D$_2$). For Bob's measurement, we fix $B_0=\sigma_z$ and $B_1=\sigma_x$. The latter can be realized by two HWPs (H$_2$ at $22.5^\circ$ and H$_{2'}$ at $0$), a PBS and two APDs. Whereas, the former can be realized by similar setup by removing the two HWPs. Two outcomes of $B_i$ are directly read by APDs (D$_3$ and D$_4$).

For the photon detection, we  register the coincidence rates between APDs of Alice and Bob. For each measurement, we record clicks for $100$s and total coincidence counts are about $1,5000$. For 1sDIST of pure two-qubit entangled states, we calculate the violation of the FGSI based on the experimental results of conditional probabilities $P(0_{B_i}|0_{A_i})$ in Fig.~\ref{fig:FGI}. By fixing $B_0$ and $B_1$ and choosing over $A_0$ and $A_1$, and the proper coefficient $\theta_\text{max}$ in $A_1$, we find that for any pure two-qubit entangled states in Eq.~(\ref{eq:state}), the maximum violation $S_\text{FGSI}=2$ of the FGSI can always be achieved. (Further details of self-testing of a general pure two-qubit entangled state are provided in Supplemental Materials~\cite{supp}).

We next identify which particular pure two-qubit entangled state has been self-tested based on the experimental result of $E$ given by Eq.~(\ref{MP}). The measurement correlations are obtained by the coincidence counts between APDs (D$_2$, D$_3$), (D$_2$, D$_4$), (D$_1$, D$_3$), and (D$_1$, D$_4$), respectively. The quantity $2E-1$ can be used to determine the concurrence of the self-tested pure state from the measurement data that gives rise to the maximal violation of the FGSI. Further, we check the monotonic relation between the value of $E$ and the concurrence $C$ as calculated by the density matrices of the states which are reconstructed via quantum state tomography,  as shown in Fig.~\ref{fig:figs}(a).

In Fig.~\ref{fig:figs}(b), we  show the measurement correlation $p(00|A_0B_0)$ as a function of the state coefficient $a$ as determined from the quantum state tomography for the four types of pure two-qubit entangled states. The monotonic relationship between $p(00|A_0B_0)$ and $a$ implies that together with self-testing via the maximal violation of the FGSI, with measurement correlations we can obtain the full knowledge of a pure two-qubit entangled state  without the requirement of quantum state tomography. Furthermore, we can identify Alice's measurement operators in our 1sDIST scenario as well.

Then we compare the state $\ket{\phi_\pm}$ ($\ket{\psi_\pm}$) with the measured coefficient $a$ obtained via the procedures of self-testing and identification compared to that obtained via quantum state tomography for each initial state being self-tested. In Fig.~\ref{fig:figs}(c), we show the fidelity between the states $\rho_\text{test}$ and $\rho_\text{tomo}$ which are obtained with two different methods
\begin{equation}
F(\rho_\text{test},\rho_\text{tomo})=\text{Tr}\sqrt{\sqrt{\rho_\text{test}}\rho_\text{tomo}\sqrt{\rho_\text{test}}},
\label{eq:fidelity}
\end{equation}
where $\rho_\text{test}=\ket{\varphi_\pm}\bra{\varphi_\pm}$ ($\varphi=\phi,\psi$) is a pure state with the coefficient $a$ obtained via the procedures of self-test and identification, and $\rho_\text{tomo}$ is the state reconstructed via quantum state tomography. The self-testing precision can be quantified by the fidelity in Eq.~(\ref{eq:fidelity}). For each experimentally generated state which is not perfect, the state purity is smaller than $1$. In our experiment, the lowest purity of the states being self-tested is about $0.950$. The lowest fidelity is $0.983$ indicating a superior precision.

%In \cite{GBD+18}, another scenario of one-sided device-independent self-testing any pure two-qubit entangled state has been proposed. Together with the maximum violation of FGSI, from the violation of Cavalcanti-Foster-Fuwa-Wiseman (CFFW) steering inequality~\cite{CFFW15}
%\begin{align}
%&S_\text{CFFW}=\sqrt{\langle(A_0+A_1)B_0\rangle^2+\langle (A_0+A_1)B_1\rangle^2}\nonumber\\&+\sqrt{\langle(A_0-A_1)B_0\rangle^2+\langle (A_0-A_1)B_1\rangle^2}
%\label{eq:CFFW}
%\end{align}
%with $\langle A_iB_j\rangle=\sum_{a,b}(-1)^{a\oplus b}p(ab|ij)$, one can particularly identify which pure two-qubit entangled state has been self-tested. By fixing the measurements of Alice and Bob as $A_0=\sigma_z$, $A_1=\cos2\theta^+_\text{max}\sigma_z+\sin2\theta^+_\text{max}\sigma_x$, $B_0=\sigma_z$ and $B_1=\sigma_x$, we obtain the experimental violation of CFFW inequality via the measured correlations. In Fig.~\ref{fig:figs}(c), the relation between the value of $S_\text{CFFW}$ and the concurrence is monotonic. However, the relation is same for the four different types of pure two-qubit entangled states~(\ref{eq:state}). Hence one can not identify which pure two-qubit entangled state has been self-tested unless the particular type of pure two-qubit entangled states is known.

{\it Robustness:}---In order to quantify the self-testing statement, we adopt the approach in~\cite{K16} to provide the robustness bound of our 1sDIST protocol given in~\cite{supp},
$Q_{\Psi,S_\text{FGSI}}:=\inf_{\rho_{AB}\in \rho(S_\text{FGSI})}\Xi(\rho_{AB}\rightarrow\Psi_{A'B'})$, where $\rho(S_\text{FGSI})$ is the set of bipartite states which violate the FGSI at least with a value $S_\text{FGSI}$, $\Xi(\rho_{AB}\rightarrow\Psi_{A'B'}):=\max_{\Lambda_A}F((\Lambda_A\otimes \one_B)\rho_{AB},\Psi_{A'B'})$ is the extractability of the test state $\rho_{AB}$ to a target state $\Psi_{A'B'}$ with the maximum taken over all quantum channels (completely positive trace-preserving maps) of the correct input/output dimension acting only on Alice's side, and $F$ is the fidelity defined in Eq.~(\ref{eq:fidelity}). The robustness can be described by the lowest possible extractability. We find that for any pure two-qubit entangled state (\ref{eq:state}), the following self-testing bound
\begin{equation}
Q_{\Psi,S_\text{FGSI}} \leq Q = a^2+(1-a^2)\frac{S_\text{FGSI}-S_\text{LHS}}{2-S_\text{LHS}},
\label{eq:Q1}
\end{equation}
with $1/\sqrt{2}\leq a <1$, holds, where $S_\text{LHS}=1+1/\sqrt{2}$ is the local hidden state bound of the FGSI~\cite{GBD+18,K16}. We prove that $Q_{\Psi,S_\text{FGSI}}= 1/2+1/2
(S_\text{FGSI}-S_\text{LHS})/(2-S_\text{LHS})$ in~\cite{supp}. We compare the fidelity $F$~(\ref{eq:fidelity}) between the self-tested state and the tomographically reconstructed state with the robustness bound $Q$~(\ref{eq:Q1}). In Fig.~\ref{fig:robustness}, for four types of pure two-qubit entangled states, the robustness of our 1sDIST protocol is ensured by the fidelity exceeding the self-testing bound for a given violation $S_\text{FGSI}$.

Finally, it may be relevant to mention that 1sDIST certifications require no-signalling constraints on the devices, which can be tested through the influence on Bob's side from the measurements of Alice's side~\cite{P13}. Using our experimental data we verify that the local marginal probabilities on Bob's side are independent of Alice's settings. From our experimental data, it can be seen that for any value of $\{\beta,B_j\}$, $\sum_{\alpha=0,1} p(\alpha\beta|A_iB_j)$ is identical within $6$ standard deviations for $i=0,1$~\cite{supp}.

%\begin{figure}%[htbp]
%\includegraphics[width=0.45\textwidth]{newfigure4}
%\caption{}
%\label{fig:probability}
%\end{figure}

%\begin{figure}%[htbp]
%\includegraphics[width=0.45\textwidth]{figure5}
%\caption{}
%\label{fig:CFFW}
%\end{figure}

{\it Conclusions:}---To summarize, in this work, we experimentally demonstrate 1sDIST of any pure two-qubit entangled state. Our experimental proof is based on the fact that a family of extremal steerable correlations can be used to self-test any pure two-qubit entangled state. Correlation functions such as mutual predictability and measurement correlation together with the maximum violation of the FGSI can  predict which particular pure two-qubit entangled state has been self-tested, as well as determine the measurement settings employed by the untrusted party. The robustness of our protocol is analytically derived through an operator inequality, and is subsequently experimentally demonstrated by the fidelity between the self-tested state and the tomographically reconstructed state exceeding the robustness bound.

Our 1sDIST protocol of any pure two-qubit entangled state is practically efficient compared to the experimentally more demanding fully device-independent self-testing~\cite{ZCP+18,ZCP+19,ZCY+19,GMM+19,GPL+19} based on the tilted Bell-CHSH inequality~\cite{YN13} for two important reasons. First, the approach based on the FGSI requires observing joint probabilities $p(\alpha\beta|A_iB_j)$ for the two pairs of observables $A_0B_0$ and $A_1B_1$, whereas, the approach based on the tilted Bell-CHSH inequality requires observing joint probabilities $p(\alpha\beta|A_iB_j)$ for all the four pairs of observables $A_0B_0$, $A_0B_1$, $A_1B_0$ and $A_1B_1$. Secondly, the latter also requires high detection efficiency in order to close the detection loophole for demonstrating Bell violation. In context of the two settings scenario, it has been shown that detection efficiency required is as high as $83\%$~\cite{GM87}. However, the 1sDIST scenario is based on violation of quantum steering inequalities, where the fair sampling condition needs to be invoked only on the untrusted side~\cite{OBSS19}. In this case it may be noted that in the two-settings scenario a detection efficiency of only $50\%$  suffices to close the detection loophole~\cite{PCSA15,SGMP16}. Thus, our approach of entanglement certification requires measurement of not only a lesser number of joint observables, but also lesser detector efficiency compared to the fully device-independent protocol, thereby promising better applicability in practical information processing scenarios.

\acknowledgments
This work has been supported by the National Natural Science Foundation of China (Grant Nos. 11674056 and U1930402) and the startup fund from Beijing Computational Science Research Centre. ZHB acknowledges support from the Natural Science Foundation of Jiangsu Province (Grant No. BK20190577) and the Fundamental Research Funds for the Central Universities (JUSRP11947). LX acknowledges support from Postgraduate Research \& Practice Innovation Program of Jiangsu Provice (KYCX18\_0056). CJ thanks Yeong-Cherng Liang, Ivan Supic and Jedrzej Kaniewski for discussions and acknowledges support from Ministry of Science and Technology of Taiwan(108-2811-M-006-501). ASM acknowledges support from the Project No. DST/ICPS/Qust/2018/98 of the Department of Science and Technology, India.

\bibliographystyle{plain}
\bibliographystyle{apsrev4-1}
\bibliography{selftest.bib}

\begin{widetext}
\renewcommand{\thesection}{\Alph{section}}
\renewcommand{\thefigure}{S\arabic{figure}}
\renewcommand{\thetable}{S\Roman{table}}
\setcounter{figure}{0}
\renewcommand{\theequation}{S\arabic{equation}}
\setcounter{equation}{0}

\section{Supplemental Materials for ``Experimental demonstration of one-sided device-independent self-testing of any pure two-qubit entangled state''}

In Supplemental Materials, we theoretically provide the robustness bound of our one-sided device-independent self-testing protocol. We also provide the experimental details for both no-signaling tests and self-testing of a family of pure two-qubit entangled states with relevant phase.

\subsection{Self-testing bound}

To obtain a robust self-testing statement in our one-sided device-independent protocol, we follow the approach of Kaniewski in~\cite{K16} which involves deriving a non-trivial lower bound on a quantity called the \emph{extractability}. In our self-testing scenario, the \emph{extractability} of the test state $\rho_{AB}$ to a target state $\Psi_{A'B'}$  is defined as
\begin{equation}
\label{extractability}
{\Xi(\rho_{AB}\to\Psi_{A'B'})}:=\max\limits_{\Lambda_A} F((\Lambda_A\otimes \openone_B) \rho_{AB},\Psi_{A'B'}),
\end{equation}
where the maximum is taken over all quantum channels (completely positive trace-preserving maps) acting on Alice's Hilbert space of the correct input/output dimension and the fidelity is defined as $F(\rho,\sigma)=\|\sqrt\rho\sqrt\sigma\|_1^2$  ($\|\cdots\|_1^2$ represents the trace form). The robustness can be described by the lowest possible extractability when one observes the violation of (at least) $S_\text{FGSI}$ on the fine-grained steering inequality (FGSI)
\begin{equation} \label{FGSI}
S_\text{FGSI}=p(\beta_{B_0}|\alpha_{A_0})+p(\beta_{B_1}|\alpha_{A_1}) \leq 1+\frac{1}{\sqrt{2}},
\end{equation}
where $p(\beta_{B_j}|\alpha_{A_i})=\frac{p(\alpha\beta|A_iB_j)}{p(\alpha|A_i)}$,
and this quantity can be captured by a function defined as
\begin{equation}
\label{self-testing bounds}
Q_{\Psi,S_\text{FGSI}}:=\inf\limits_{\rho_{AB}\in \rho(S_\text{FGSI})} \Xi(\rho_{AB}\to\Psi_{A'B'}),
\end{equation}
where $\rho(S_\text{FGSI})$ is the set of bipartite states which violate FGSI at least with a value $S_\text{FGSI}$.

Let us now consider the case when the target state is the maximally entangled state.
If the FGSI is not violated, we cannot improve over the trivial lower bound of $1/2$ i.e.~$Q_{\Psi,S_\text{FGSI}}(S_\text{LHS}) = 1/2$, where $S_\text{LHS}$ is the LHS bound of the FGSI. On the other extreme, by assumption we have $Q_{\Psi,S_\text{FGSI}}(S_{Q})=1$, where $S_{Q}=2$. Note that every intermediate violation can be achieved as a mixture of these two points, for instance, a mixture of the product state $\rho_{A} \otimes \rho_B$, where $\rho_A=\openone/2$ and $\rho_B$ is one of the eigenstates of the Pauli observables $(\sigma_z \pm \sigma_x)/\sqrt{2}$, and the maximally  entangled state  $\left( \ket{00}+  \ket{11}\right)/\sqrt{2}$ can give rise to any intermediate violation for the Pauli observables $A_0=\sigma_z,A_1=\sigma_x$ on Alice's side  and the Pauli observables $B_0=\sigma_z$, $B_1=\sigma_x$ on Bob's side. This leads to an upper bound of the form
\begin{equation}
\label{eq:upper-bound}
Q_{\Psi,S_\text{FGSI}} \leq \frac{1}{2} + \frac{1}{2} \cdot \frac{S_\text{FGSI} - S_\text{LHS}}{S_{Q} - S_\text{LHS}}.
\end{equation}

We now adopt the operator inequality method given in Ref. \cite{K16} which has been used to obtain a nontrivial lower bound on the extractability and demonstrate that it equals  the upper bound given by Eq. (\ref{eq:upper-bound}). Since the FGSI is not linear in the possible assemblage $\{\sigma_{\alpha|A_i}=p(\alpha|A_i)\rho_{\alpha|A_i}\}$ realizing the given violation, for the purpose of applying the operator inequality method we approximate the FGSI as the linear inequality in the following. Suppose Bob fixes his observables of measurements as $B_0=\sigma_z$ and $B_1=\sigma_x$ and Alice chooses her measurements $A_0$ and $A_1$. Then the correlation between systems A and B violates the FGSI maximally if and only if the bipartite state is the  pure two-qubit entangled state, for instance
\begin{align}
\label{eq:state}
&\ket{\phi_+}=a\ket{00}+\sqrt{1-a^2}\ket{11}, \frac{1}{\sqrt{2}} \le a <1,
\end{align}
and the choices of Alice's observables  are given by $\{A_0=\sigma_z,A_1=(1-2a^2)\sigma_z + 2a\sqrt{1-a^2}\sigma_x\}$ up to local isometry. Note that when the above quantum state and the measurements realize the maximal violation of the FGSI, the marginal probabilities on Alice's side for the outcome $0$ are given by $p(0|A_0)=a^2$ and $p(0|A_1)=\left[1-(1-2a^2)^2\right]/2$. We now linearize the FGSI as follows:
\begin{equation}\label{LSI}
\frac{p(00|00)}{a^2}+\frac{2p(00|11)}{1-(1-2a^2)^2} \le
1+\frac{1}{\sqrt{2}}.
\end{equation}
It can be checked that the above linear steering inequality is maximally violated by any pure entangled state (\ref{eq:state}).

Let us now apply the operator inequality method to the  linearized steering inequality (\ref{LSI}) in the case of the maximally entangled state for  the steering scenario where Bob performs the following Pauli measurements: $B_0=\sigma_x$ and $B_1=\sigma_z$. In this case let our target be of the form
\begin{equation}
\targetstate=\frac{1}{4} \left[\one_A \otimes \one_B + \sigma_y \otimes \sigma_y +\frac{1}{\sqrt{2}}\left(\sigma_x \otimes \sigma_x + \sigma_x \otimes \sigma_z+\sigma_z \otimes \sigma_x- \sigma_z \otimes \sigma_z\right) \right].
\end{equation}
It can be checked that if Alice performs measurements in the basis of the Pauli observables $A_0=\left(\sigma_x+\sigma_z\right)/\sqrt{2}$ and $A_1=\left(\sigma_x-\sigma_z\right)/\sqrt{2}$, the above state violates our steering inequality
\begin{equation}
2p(00|00)+2p(00|11) \le  1+\frac{1}{\sqrt{2}}
\end{equation}
maximally.

We consider the extraction channel of the form
\begin{equation}
 [\Lambda_A(\vartheta)](\cdot) =\frac{1+g(\vartheta)}{2} (\cdot)+\frac{1+g(\vartheta)}{2} \Gamma(\vartheta) (\cdot)  \Gamma(\vartheta),
\end{equation}
where $g(\vartheta)=(1+\sqrt{2})(\sin \vartheta +\cos \vartheta-1)$ and
\begin{align}
\Gamma(\vartheta)&=\left\{
\begin{array}{lr}
\sigma_x &  \quad  \text{for} \quad  \vartheta \in \left[0, \frac{\pi}{4}\right]   \\
\sigma_z  &  \quad  \text{for} \quad   \vartheta \in \left(\frac{\pi}{4}, \frac{\pi}{2}\right] \\
\end{array}
\right\},
\end{align}
which has been used to obtain analytic self-testing bound for the Clauser-Horne-Shimony-Holt (CHSH) Bell inequality in~\cite{K16}. Our goal is to prove an operator inequality of the form
\begin{equation}
\label{eq:operator-inequality}
K(\vartheta) \geq s W(\vartheta) + \mu \one
\end{equation}
for all $\vartheta$ and suitably chosen (real) constants $s$ and $\mu$. Here $K$ is the dephasing operator in the Hadamard basis given by
\begin{align}
  K(\vartheta)&=(\Lambda_A(\vartheta) \otimes \openone_B)   (\targetstate) \nonumber \\
      &= \frac{1}{4} \left[\one_A \otimes \one_B + g(\vartheta) \sigma_y \otimes \sigma_y +\frac{1}{\sqrt{2}}\left(\sigma_x \otimes \sigma_x +  \sigma_x \otimes \sigma_z+g(\vartheta) \sigma_z \otimes \sigma_x- g(\vartheta) \sigma_z \otimes \sigma_z\right) \right]
\end{align}
and $W$ is the steering operator in which Alice's observables are rotated by an angle $\vartheta$ which is given by
\begin{align}
% W(\vartheta)&=\frac{1}{4}\big[\one_A \otimes \one_B +\cos \vartheta \sigma_x \otimes \one_B+\frac{1}{2} \left(\one \otimes \sigma_x + \one \otimes \sigma_z  \right)  \nonumber \\
 %&+\frac{1}{2}\left(\cos \vartheta \sigma_x \otimes \sigma_x+  \sin \vartheta \sigma_z \otimes \sigma_x +  \cos \vartheta \sigma_x \otimes \sigma_z -  \sin \vartheta \sigma_z \otimes \sigma_z  \right) \big] \\
 W(\vartheta)&=\one_A \otimes \one_B +\cos \vartheta \sigma_x \otimes \one_B+\frac{1}{2} \left(\one \otimes \sigma_x + \one \otimes \sigma_z  +\cos \vartheta \sigma_x \otimes \sigma_x+  \sin \vartheta \sigma_z \otimes \sigma_x +  \cos \vartheta \sigma_x \otimes \sigma_z -  \sin \vartheta \sigma_z \otimes \sigma_z  \right).
\end{align}

We now check whether the operator
\begin{align}
% T(\vartheta)&=K(\vartheta)-s W(\vartheta) -\mu  \one_{AB} \nonumber \\
 %&=\frac{1}{4} \Bigg\{ (1-s - 4 \mu) \one \otimes \one -s \left[\cos \vartheta \sigma_x \otimes \one -\frac{1}{2}\left(\one \otimes \sigma_x+ \one \otimes \sigma_z  \right) \right]
 %+g(\vartheta) \sigma_y \otimes \sigma_y \nonumber \\
 %&+\frac{1}{\sqrt{2}}\left[\left(1-\frac{s\cos\vartheta }{\sqrt{2}}\right) \sigma_x \otimes \sigma_x + \left(1-\frac{s\cos %\vartheta}{\sqrt{2}}\right) \sigma_x \otimes \sigma_z
 %+ \left(g(\vartheta)-\frac{s\sin \vartheta}{\sqrt{2}}\right) \sigma_z \otimes \sigma_x- \left(g(\vartheta)-\frac{s\sin \vartheta}{\sqrt{2}}\right) \sigma_z \otimes \sigma_z\right]\Bigg\} \\
  T(\vartheta)&=K(\vartheta)-s W(\vartheta) -\mu  \one_{AB} \nonumber \\
 &= \left(\frac{1}{4}-s -  \mu\right) \one \otimes \one -s \left[\cos \vartheta \sigma_x \otimes \one -\frac{1}{2}\left(\one \otimes \sigma_x+ \one \otimes \sigma_z  \right) \right]
 +\frac{g(\vartheta)}{4} \sigma_y \otimes \sigma_y \nonumber \\
 &+\left(\frac{1}{4\sqrt{2}}-\frac{s\cos \vartheta}{2}\right) \left(\sigma_x \otimes \sigma_x +  \sigma_x \otimes \sigma_z \right)
 + \left(\frac{g(\vartheta)}{4\sqrt{2}}-\frac{s\sin \vartheta}{2}\right) \left(\sigma_z \otimes \sigma_x-  \sigma_z \otimes \sigma_z \right)
\end{align}
is positive semi-definite for
\begin{equation}
 s=\frac{\sqrt{2}}{2(\sqrt{2}-1)}; \quad  \mu=-\frac{1}{\sqrt{2}-1}.
\end{equation}
It suffices to consider the range $\vartheta \in [0, \frac{\pi}{4}]$ as in the case of the CHSH inequality \cite{K16} to check  whether $T(\vartheta)$ is positive semi-definite.

Noticing that $\left[T(\vartheta),\sigma_y \otimes \sigma_y \right]=0$ for all $\vartheta \in \left[0,\pi/2\right]$ leads us to consider the projectors
\begin{equation}
P_x=\frac{1}{2}\left[\one \otimes \one + (-1)^x \sigma_y \otimes \sigma_y \right]
\end{equation}
for $x \in \{0,1\}$. Then
\begin{equation}
 \tr M_x%=\tr P_x T(\vartheta)= \frac{1}{2}\left[1-s-4 \mu +(-1)^x g(\vartheta) \right]
 =2\left[\frac{1}{4}-s- \mu +(-1)^x \frac{g(\vartheta)}{4} \right]
\end{equation}
and
\begin{align}
 \tr M^2_x%&=\tr P_x \left[T(\vartheta)\right]^2= \frac{1}{8}\left[(1-s-4 \mu)^2+s^2(\cos^2\vartheta+\frac{1}{2})+g(\vartheta)^2+\left(1-\frac{s\cos \vartheta}{\sqrt{2}}\right)^2
 %+\left(g(\vartheta)-\frac{s\sin \vartheta}{\sqrt{2}}\right)^2\right] \nonumber \\
 %&+\frac{1}{2}(-1)^x\left(1-\frac{s\cos \vartheta}{\sqrt{2}}\right)\left(g(\vartheta)-\frac{s\sin \vartheta}{\sqrt{2}}\right)
 &=2 \left\{\left[\frac{1}{4}-s -  \mu \right]^2+s^2(\cos^2\vartheta+\frac{1}{2})+\frac{g(\vartheta)^2}{16}+\left(\frac{1}{4}-\frac{s\cos \vartheta}{\sqrt{2}}\right)^2
 +\left(\frac{g(\vartheta)}{4}-\frac{s\sin \vartheta}{\sqrt{2}}\right)^2\right\} \nonumber \\
 &+(-1)^x2\left\{\frac{g(\vartheta)}{4}\left(\frac{1}{4}-s -  \mu\right) + \left(\frac{1}{4}-\frac{s\cos \vartheta}{\sqrt{2}}\right) \left(\frac{g(\vartheta)}{4}-\frac{s\sin \vartheta}{\sqrt{2}}\right)\right\}.
\end{align}
The positivity of $\lambda_x(\vartheta):=(\tr M_x)^2-\tr M^2_x$ follows, as it can be checked that
$\lambda_x(\vartheta)>0$ for all $\vartheta \in \left[0, \pi/4\right]$ and $x$. Therefore, we obtain the following lower bound
on the extractability:
\begin{equation}
 \tr \left( K(\vartheta) \targetstate \right) = {\Xi(\rho_{AB}\to\Psi_{A'B'})} \geq s \tr \left( W(\vartheta) \targetstate \right) + \mu \tr \left(  \targetstate \right)= \frac{\sqrt{2}}{2(\sqrt{2}-1)}S_\text{FGSI} - \frac{1}{\sqrt{2}-1},
\end{equation}
which saturates the upper bound given by Eq. (\ref{eq:upper-bound}). Therefore, we have the following self-testing statement:
\begin{equation} \label{STBME}
Q_{\Psi,S_\text{FGSI}}  = \frac{1}{2} + \frac{1}{2} \cdot \frac{S_\text{FGSI} - S_\text{LHS}}{S_{Q} - S_\text{LHS}}
\end{equation}
with $S_\text{LHS}=1+1/\sqrt{2}$ and $S_{Q}=2$.

In case our target state is  any pure two-qubit entangled state,
the upper bound on $Q_{\Psi,S_\text{FGSI}}$ is given by
\begin{equation}
Q_{\Psi,S_\text{FGSI}} \le \lambda_\text{max}^2 + (1-\lambda_\text{max}^2) \cdot \frac{S_\text{FGSI}-S_\text{LHS}}{S_{Q}-S_\text{LHS}},
\end{equation}
where $\lambda_\text{max}$ is the largest Schmidt coefficient of our target state.
The above bound may be taken as the self-testing statement for any pure entangled state in our
1sDIST scenario based on the fine-grained steering inequality.

\begin{figure}
\includegraphics[width=0.5\textwidth]{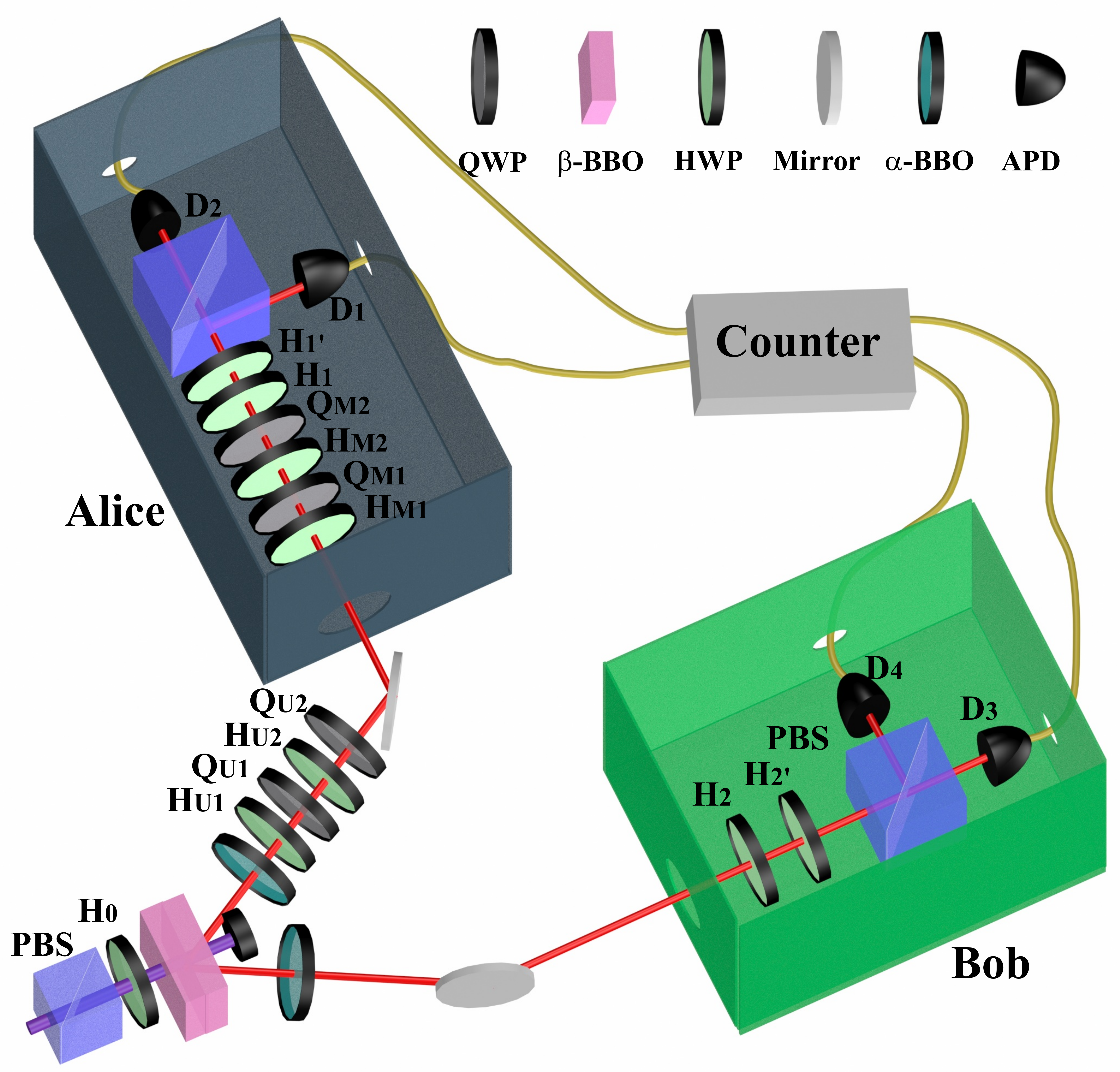}
\caption{Experimental setup for one-sided device-independent self-testing a pure two-qubit entangled state $\ket{\phi_\delta}$.
}
\label{fig:setup2}
\end{figure}

\begin{figure}
\includegraphics[width=0.4\textwidth]{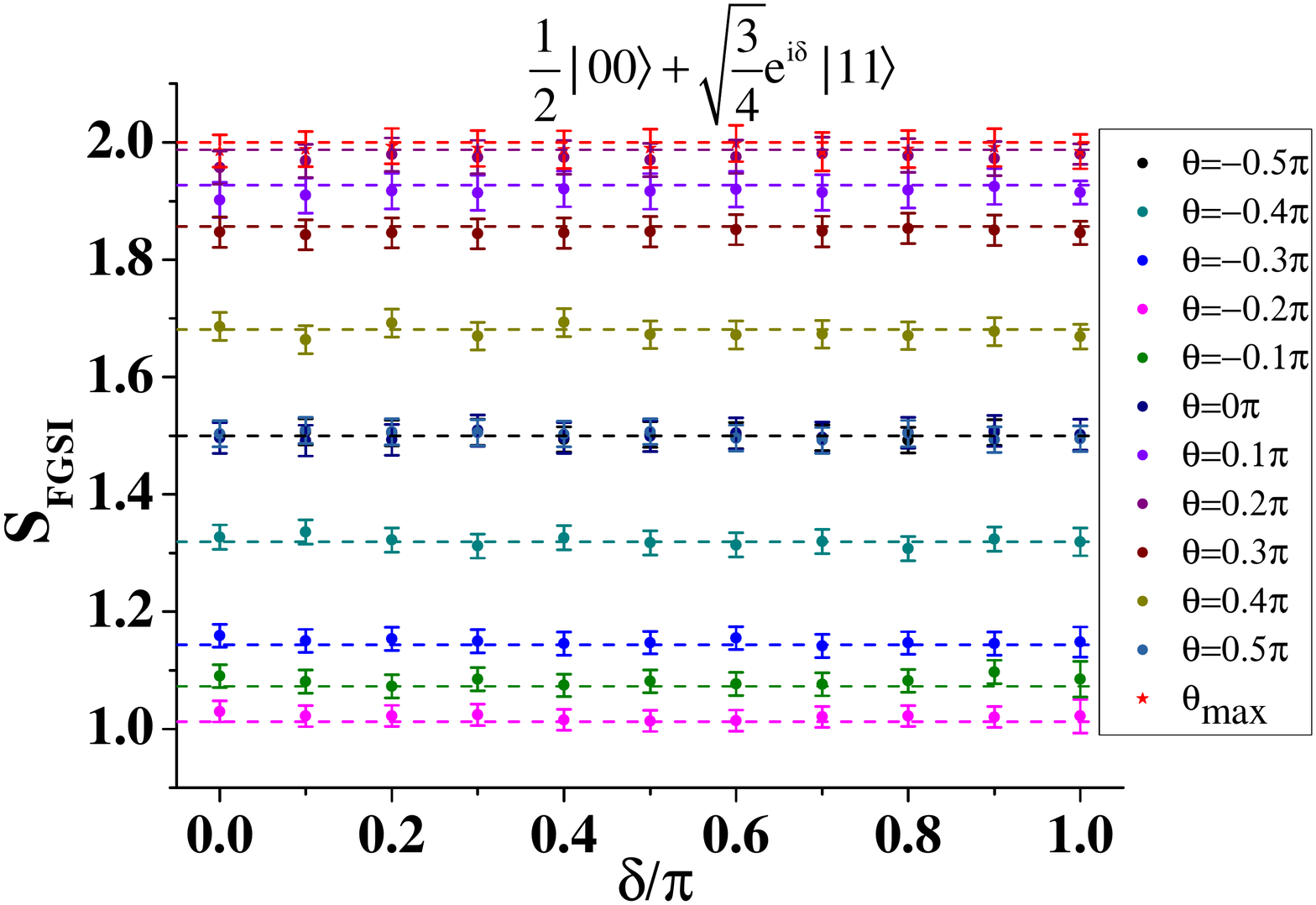}
\caption{Experimental results of $S_\text{FGSI}$ as a function of the phase $\delta/\pi$ for the state $\ket{\phi_\delta}$. Different symbols represent different measurements $\{\tilde{A}_0, \tilde{A}_1\}$ for Alice. Dashed lines represent the theoretical predictions.}
\label{fig:phase}
\end{figure}

\begin{figure*}%[b]
\includegraphics[width=0.8\textwidth]{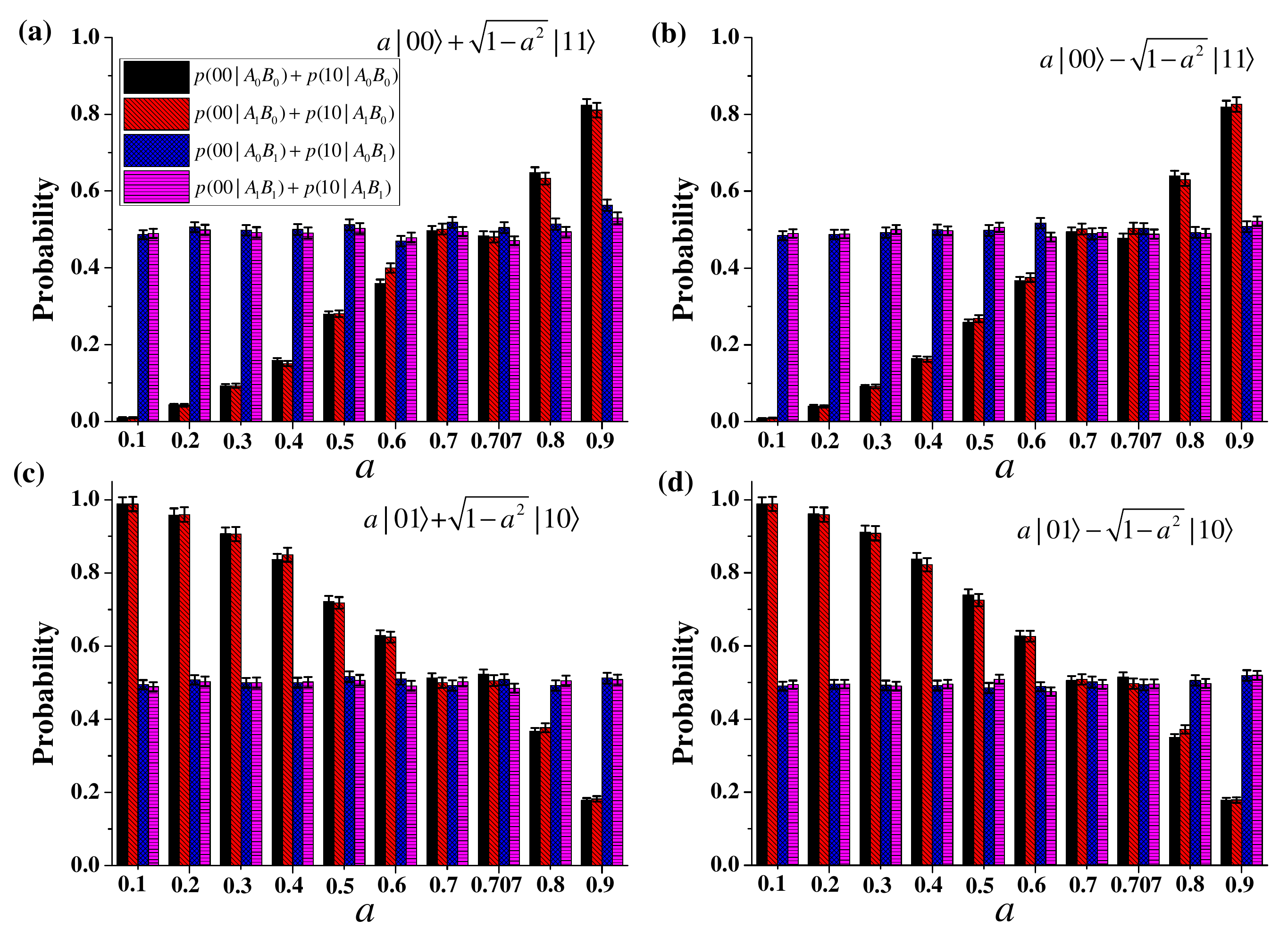}
\caption{No-signaling test for Alice's devices via the influence of Alice's measurement on Bob's side. We compare sum of measurement correlations $\{\sum_{\alpha=0}^1p(\alpha0|A_0B_0),\sum_{\alpha=0}^1p(\alpha0|A_1B_0)\}$ and $\{\sum_{\alpha=0}^1p(\alpha0|A_0B_1),\sum_{\alpha=0}^1p(\alpha0|A_1B_1)\}$.
}
\label{fig:nosignal}
\end{figure*}

\subsection{Experimental set-up for self-testing of $\ket{\phi_\delta}$}

In the main text, we consider pure two-qubit entangled states with the form given by Eq.~(2) as any pure two-qubit entangled state can always be written in the form of~(2) following Schmidt decomposition. Entangled photons are generated via type-I spontaneous parametric down-conversion (SPDC), where two joint $0.5$mm-thick-$\beta$-barium-borate ($\beta-$BBO) crystals are pumped by a continuous-wave diode laser. The visibility of entangled photonic state is larger than $0.950$. The heralded single photons pass through a polarizing beam splitter (PBS) followed by a half-wave plate (HWP, H$_0$) with the certain setting angle $\chi$ and the matrix from of the operation of the HWP
$\begin{pmatrix}
    \cos 2\chi & \sin 2\chi \\
    \sin 2\chi & -\cos 2\chi \\
    \end{pmatrix}$. The matrix form of the operation of the QWP is $\begin{pmatrix} \cos^2 \chi+i \sin^2 \chi & (1-i)\sin\chi\cos\chi\\ (1-i)\sin\chi\cos\chi & \sin^2\chi+i\cos^2\chi \\ \end{pmatrix}$.

Now we choose a family of pure two-qubit entangled states with relevant phase $\ket{\phi_\delta}=\frac{1}{2}\ket{HH}+\sqrt{\frac{3}{4}}e^{i\delta}\ket{VV}$ as an example to show it can be self-tested via our one-sided device-independent scenario. The detailed experimental setup for self-testing of a pure two-qubit entangled state $\ket{\phi_\delta}$ is shown in Fig.~\ref{fig:setup2}. By inserting a set of wave plates---HWP (H$_{U1}$ at $0$)-QWP (Q$_{U1}$ at $45^\circ$)-HWP (H$_{U2}$ at $\frac{\delta}{4}$)-QWP (Q$_{U2}$ at $45^\circ$) into one of the optical paths right after $\alpha$-BBO crystal, photon pairs are prepared into the entangled state $\ket{\phi_\delta}=\frac{1}{2}\ket{HH}+\sqrt{\frac{3}{4}}e^{i\delta}\ket{VV}$. By tuning the setting angle of the second HWP of the set of wave plates, we can prepare a family of entangled states with different $\delta$. Then with HWPs and QWPs, the state is transformed into $\ket{\phi_{\delta=0}}$ for self-testing via the maximum violation of the FGSI. A pure two-qubit entangled state $\ket{\phi_\delta}$ can be transformed into the state with the form in Eq.~(2), for example $\ket{\phi_{\delta=0}}$ via local operations, i.e., $\ket{\phi_{\delta=0}}=U_\delta\otimes\one=\begin{pmatrix}
    1 & 0 \\
   0 & e^{-i\delta} \\
    \end{pmatrix}\otimes \begin{pmatrix}
    1 & 0 \\
   0 & 1 \\
    \end{pmatrix}\ket{\phi_\delta}$. In our experiment, the local operation $U_\delta$ can be realized by a set of wave plates---HWP ((H$_{M1}$ at $0$)-QWP (Q$_{M1}$ at $-45^\circ$)-HWP (H$_{M2}$ at $\frac{\delta}{4}$)-QWP (Q$_{M2}$ at $-45^\circ$) applied on Alice's photon. Then the state can be self-tested by the maximum violation of the FGSI with the measurements for Alice and Bob $\{A_0=\sigma_z, A_1=\cos2\theta_\text{max}\sigma_z+\sin2\theta_\text{max}\sigma_x\}$ [$\theta_\text{max}=\frac{1}{2}\arccos(1-2a^2)$], and $\{B_0=\sigma_z, B_1=\sigma_x\}$. Thus, up to local isometry on Alice's side the maximum violation of the FGSI certifies any pure two-qubit entangled state in our one-sided device-independent scenario.

As shown in Fig.~\ref{fig:phase}, we find that by fixing $B_0=\sigma_z$, $B_1=\sigma_x$, and scan the parameter $\theta$ of Alice's measurements $\tilde{A}_0=\begin{pmatrix}
    1 & 0 \\
   0 & e^{-i\delta} \\
    \end{pmatrix}$ and $\theta$ of $\tilde{A}_1=\begin{pmatrix}
    \cos2\theta & e^{-i\delta}\sin2\theta \\
   \sin2\theta & -e^{-i\delta}\cos2\theta \\
    \end{pmatrix}$, for each $\delta$ one can find the certain parameter $\theta_\text{max}$ of $\tilde{A}_1$ to achieve the maximum violation $S_\text{FGSI}=2$ of the FGSI and $\theta_\text{max}$ is the same as the one for the state $\ket{\phi_{\delta=0}}=\frac{1}{2}\ket{HH}+\sqrt{\frac{3}{4}}\ket{VV}$ to achieve the maximum violation $S_\text{FGSI}=2$ of the FGSI. Thus, up to local isometry on Alice's side the maximum violation of the FGSI certifies any pure two-qubit entangled state in our one-sided device-independent scenario.

\subsection{No-signaling tests}

One-sided device-independent certifications require no-signaling constraints on the devices, which can be tested through the influence on Bob's side from the measurements of Alice's side. No-signaling constraints in our experiment expressed as
\begin{align}
\sum_{\alpha=0}^1p(\alpha\beta|A_0B_0)=\sum_{\alpha=0}^1p(\alpha\beta|A_1B_0),\text{  }
\sum_{\alpha=0}^1p(\alpha\beta|A_0B_1)=\sum_{\alpha=0}^1p(\alpha\beta|A_1B_1).
\label{S1}
\end{align}
These constrains have a clear physical interpretation: they imply that the local marginal probabilities of Bob $p(\beta|B)\equiv p(\beta|A_iB_j)=\sum_{\alpha=0}^1p(\alpha\beta|A_iB_j)$ are independent of Alice's measurement setting $A_i$, and thus Alice can not signal to Bob by her choice of input.

As in our experiment the maximal violation of the FGSI corresponds to $\alpha=\beta=0$, we show the local marginal probabilities of Bob $p(0|A_0B_0)$, $p(0|A_1B_0)$, $p(0|A_0B_1)$ and $p(0|A_1B_1)$ with $A_0=\sigma_z$, $A_1=(1-2a^2)\sigma_z+2a\sqrt{1-a^2}\sigma_x$, $B_0=\sigma_z$ and $B_1=\sigma_x$ in Fig.~\ref{fig:nosignal}. One can see that $p(0|A_0B_0)$ and $p(0|A_1B_0)$, $p(0|A_0B_1)$ and $p(0|A_1B_1)$ are approximately identical for various input states, respectively, which implies the no-signalling constraints are satisfied. Note that testing no-signalling from Alice to Bob is relevant in the context of the one-sided device-independent self-testing scheme, on the other hand, no-signalling from Bob to Alice is taken for granted~\cite{P13}. From our experimental data, it can be seen that for any value of $\{\beta,B_j\}$, $\sum_{\alpha=0,1} p(\alpha\beta|A_iB_j)$ is identical within $6$ standard deviations for $i=0,1$, which satisfies Eq.~(\ref{S1}).

\end{widetext}

\end{document}